# Improving bit-vector representation of points-to sets using class hierarchy

Hamid A. Toussi
Department of Mathematics & Computer Science
University of Sistan and Baluchestan
Zahedan, Iran
e-mail: hamid2c@gmail.com

Ahmed Khademzadeh
Department of Computer Engineering
Islamic Azad University, Mashhad Branch
Mashhad, Iran
e-mail: khademzadeh@mshdiau.ac.ir

*Abstract*—Points-to analysis is the problem of approximating run-time values of pointers statically or at compile-time. Points-to sets are used to store the approximated values of pointers during points-to analysis. Memory usage and running time limit the ability of points-to analysis to analyze large programs.

To our knowledge, works which have implemented a bit-vector representation of points-to sets so far, allocates bits for each pointer without considering pointer's type. By considering the type, we are able to allocate bits only for a subset of all abstract objects which are of compatible type with the pointer's type and as a consequence improve the memory usage and running time. To achieve this goal, we number abstract objects in a way that all the abstract objects of a type and all of its sub-types be consecutive in order.

Our most efficient implementation uses about 2.5× less memory than hybrid points-to set (default points-to set in Spark) and also improves the analysis time for sufficiently large programs.

## I. INTRODUCTION

Points-to analysis is the problem of approximating run-time values of pointers statically or at compile-time. The result of this analysis may be used for program optimization, debugging or understanding. One of the most famous and widely used algorithms to solve this problem, which we have used in our implementations, is Andersen style or inclusion based points-to analysis [1].

Points-to sets are used to store the approximated values of pointers during the analysis. Previous works [6], [2], [3], [7], [4], [5], [15] have implemented efficient and compact points-to sets which make the overall analysis more efficient. Points-to sets which are represented based on bit-vectors [7], [3], [8] and points-to sets which are represented and manipulated based on BDD relations are shown to be more efficient [6], [2], [4], [5], [9], [15]. In this paper we utilize bit-vector representaion of points-to sets.

Points-to analysis produces several points-to sets that are similar in the sense that many of their members are common [7]. Based on this fact, both BDD based and bit-vector based methods try to have more efficient representation of points-to sets by trying to share common members.

In a pure bit-vector implementation of points-to sets, one bit is allocated for every abstract object (allocation-site) that is reachable from root methods. Reachability of allocation-sites can be determined using a conservative call-graph. This call-graph can be created using either CHA [10], RTA [11] or VTA [12] methods.

Hientze have implemented a shared bit-vector representation of points-to sets in order to improve memory and time efficiency [7]. These sets consist of two parts, a shared bit-vector (base part) and an overflow list. The base part is shared among two or more points-to sets and the overflow lists is maintained to include 20 or fewer members. This representation benefits from the fact that many points-to sets are similar. Hirzel also used this implementation of points-to sets in his work [8]. Lhotak and Hendren [3] used a variant of this technique. Their set had no mechanism to share common subsets. Once any set gets larger than some specified-size (e.g. 16 members), it becomes a pure bit-vector points-to set.

In strongly typed languages like Java [14] more precise points-to results and, as a consequence, smaller points-to sets can be achieved by using type filtering during points-to set propagation. This method which is known as online type filtering [13] also makes the analysis faster [3].

Previous works, we have mentioned above, haven't made use of types in order to improve size of points-to sets, e.g., in the pure bit-vector implementation of a points-to set associated with a variable of type T, one bit is allocated for every abstract object of the program. By noticing the types, we are able to allocate bits only for a subset of all the abstract objects which are of compatible type with the variable (i.e. abstract objects of type T or its subtypes). To be able to make use of this fact, we reordered abstract objects of programs, using class-hierarchy so as to assign abstract objects of compatible types, successive numbers.

We have a number of implementations and the best of them employs the above-mentioned idea and also uses Spark's hybrid points-to set's idea to gain more efficiency.

There is also another related work, sparse bitmaps, which is employed by GCC [16] and LLVM [17] compilers. Every sparse bitmap consists of a linked list of elements and each element is a bit-vector of size eight words. No element is allocated in a sparse bitmap unless it has at least one non-zero bit set. In order to calculate the memory usage of sparse bitmaps when it is combined with our desired ordering, we did not need to implement a combined version of our method and sparse bitmaps since the memory gain can be calculated by examining bit-vectors after the propagation and finding out how many elements could be saved. This combined version could have some minor memory improvement compared to our current implementation, but considering the cost of linked list manipulation and minor memory usage improvement, we believe that this memory gain is not worth it (see section VII).

Some background information comes in section II. As stated above, we would first number the abstract objects in our desired order and then we need to implement a bit-vector (we call it **RangedBitVector**) which takes these ranges into account. An overall picture of the work is given in section III. Numbering is explained in section IV and **RangedBitVector** is described in section V. Section VI concludes our implementations and we show our experimental results in section VII.

## II. BACKGROUND

Andersen or inclusion based points-to analysis which is used in this paper, consists of a set of deduction rules. Points-to propagation is the process of applying Andersen's rules repeatedly until reaching a fix-point where all the constraints are satisfied, and as a consequence, a safe solution to the points-to analysis problem is found. This set of constraints are inferred from the input program and are represented by Pointer Assignment Graph (PAG) in Spark.

PAG consists of different nodes and edges. Pointers of the input program are represented by variable nodes. They represent local variables, method parameters and static fields which can hold pointer values. Allocation nodes are abstract objects (i.e. every allocation node abstracts a set of run-time objects) and every allocation node corresponds to an allocation-site of the input program.

Edges mainly represent flow of points-to sets. For example an edge from the variable node *v* to variable node *w* shows the statement w=v and makes the propagator to add all allocation nodes in points to set of *v* to points-to set of *w*.

There are other edges and nodes in the PAG (store edges, load edges, allocation edges, concrete field nodes and field dereference nodes). See [3], [18] for more details about them.

## III. OVERVIEW

Suppose there are 1000 allocation-sites within all the reachable methods of a program and from these, there are only 200 allocation-sites of type A (i.e. instantiation or instantiations of class A at the allocation-site), or one of its sub-types. If we have a variable *v* of declared type A, and we are supposed to allocate a bit-vector representation of points-to set for it, we can allocate 200 bits (one bit per each allocation-site of type A or its sub-types) instead of 1000 bits. To achieve this, we have to number allocation nodes so that all of the allocation nodes of a declared type, and all of its sub-types, fall within the same interval and be consecutive in order (e.g. the interval [201, 400] in the example above). Recall that for every allocation-site one allocation node is created.

Every non-interface type is mapped to an interval as described above. An interface type may be associated with more than one interval (one interval for every top subclass that have implemented the interface). Every **RangedBitVector** has an interval and is able to handle different operations.

During points-to set propagation, points-to sets for two kinds of nodes (variable nodes and concrete field nodes. see [3]) are created. For each of these nodes, a specialized bit-vector implementation of points-to set is allocated based on type of the node (i.e. declared type of variables in variable nodes and fields in concrete field nodes).
To allocate the specialized points-to set, the interval or intervals associated with the node's type are required so allocation node manager (the module we added to Spark to manipulate intervals and allocation nodes in our desired way) is queried and type of the node is sent to it as a parameter to achieve these intervals. Finally, the points-to set for the node is allocated. This points-to set has a **RangedBitVector** for every interval associated with the type. This is the description of **RangedPointsToSet**, another variant of it is also implemented in this work (**RangedHybridPointsToSet**). See sections VI and V.

## IV. NUMBERING

Originally, Spark (the framework we have used to implement our work) numbers allocation nodes as allocates them in the PAG, which is not our desired way of numbering. The procedure in Figure 1 renumbers the allocation nodes and associates every reference type with an interval. This does not include interface types which may be associated with more than one interval.
The interval's lower bound indicates the number associated with the first allocation node of that type, and its sub-types and interval's upper bound indicates the number associated with the last allocation node within the interval.

```
void dfsVisit(SootClass cl){
  lower = globalCounter + 1;
  for (AllocNode alloc : class2allocs.get(cl)){
    globalCounter = globalCounter + 1;
    globalArray[globalCounter] = alloc;
  }

  // cha is the class hierarchy
  subclasses = cha.getSubClassesOf(cl);

  if (subclasses.isEmpty()){
    // SootClass cl is a leaf in the class hierarchy
    type2interval.put(cl.getType(),
        new Interval(lower, globalCounter));
    return;
  }

  for (SootClass c : subclasses)
    dfsVisit(c);
  upper = globalCounter;
  type2interval.put(cl.getType(),
      new Interval(lower, upper));
}
```

Fig. 1. Procedure to renumber Allocation Nodes and create intervals

The procedure is basically a depth first traversal of the class hierarchy. The index of an allocation node in the `globalArray` represents its corresponding bit in a pure bit-vector implementation. In fact, the `globalArray` represents the universe of all the allocation nodes.

In this procedure two hash-maps are used. Each reference type is mapped to a linked-list of allocation nodes of that type through the map `class2allocs`. The map `type2interval` gets filled during this procedure and it maps every type to its associated interval. Note that `globalArray`, `globalCounter`, `class2allocs` and `type2interval` are global (static in term of Java). The first invocation of this procedure is `dfsVisit(getSootClass("java.lang.Object"))`. Figure 2 shows an example class hierarchy and the intervals associated with each type as a result of the invocation.

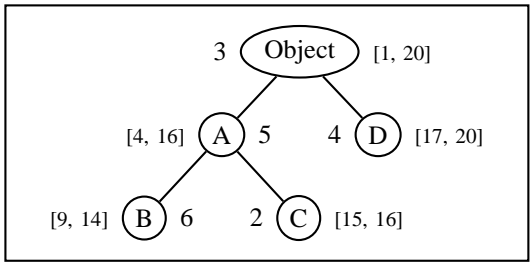

Fig. 2. Each node represents a type in the class hierarchy. The number next to each node shows the number of allocation nodes of that type in the PAG. Intervals associated with each type is created in postorder (i.e. B, C, A, D, Object) and are shown in brackets.

## V. RANGED BIT-VECTORS

During the points-to set propagation, many set union operations are performed. For example, an assignment edge from variable node $w$ to variable node $v$ make the propagator to add all the members of $pt(w)$ to $pt(v)$ if it is not already satisfied. Note that $pt(v)$ shows the points-to set of variable node $v$. Considering bit-vectors, this set union operation is implemented using logical **or** operation.
We have implemented a customized version of bit-vectors which takes the range associated with each bit-vector into account so that the logical **or** operation can be done efficiently. We have called this implementation of bit-vectors, **Ranged-BitVector**. Using the intervals not only saves memory needed for points-to sets, but also helps us to do points-to propagation more efficiently by limiting the size of bit-vectors.
The set union operation $pt(v) = pt(v) \cup pt(w)$ is actually implemented as:

```
for (RangedBitVector bv : pt(v).bitvectors)
  for (RangedBitVector bw : pt(w).bitvectors)
    or(bv, bw);
```

In previous works each bit-vector is **and**ed with a suitable type mask to enforce type filtering.
Type masks are created before the propagation and are used for type filtering during the propagation [8]. The suitable

```
boolean or(RangedBitVector x, RangedBitVector y){
  /* The lower bounds are aligned. For example,
     if we are supposed to allocate a ranged
     bitvector of length 10 for the interval
     [10, 20], assuming chunk size is 8, 2 bytes
     will be allocated and the lower bound is
     aligned to 8. */
  if (y.isSubrangeOf(x)){
    /* x is super-range and y is subrange */
    /* Absolute index of lower chunk of the
       subrange */
    L = indexOf(y.lower);
    /* Absolute index of upper chunk of the
       subrange */
    U = indexOf(y.upper);
    /* Relative index of lower chunk of
       the subrange which always is 0 */
    firstChunkY = 0;
    /* Relative index of upper chunk of the
       subrange */
    lastChunkY = U - L;
    /* Absolute index of lower chunk of the
       super-range */
    int temp = indexOf(x.lower);
    /* Relative index of first common chunk
       within super-range */
    firstChunkX = L - temp;
    /* Relative index of last common chunk
       within super-range */
    lastChunkX = firstChunkX + lastChunkY;
  } else if (x.isSubrangeOf(y)) {
    /* y is super-range and x is subrange */
    /* similar to the previous case ... */
  } else {
    return false;
  }
  ret = false;
  for (int i = firstChunkX, j = firstChunkY;
       i <= lastChunkX; i++, j++) {
    if (!ret) old = x.bits[i];
    x.bits[i] |= y.bits[j];
    if (!ret)
        if (old != x.bits[i]) ret = true;
  }
  return ret;
}

int indexOf(int index) {
    return index / CHUNK_SIZE;
}
```

Fig. 3. The procedure to **or** 2 RangedBitVectors

type mask is determined based on type of the points-to set containing the bit-vector. In fact, each type is associated with one type mask. This means, to perform on the fly type filtering [13], one **and** operation for every set union operation is needed. With our implicit type filtering mechanism this pass of points-to analysis which was needed to create type masks is eliminated and also we does not need the additional **and** operation either.

Our method performs type filtering implicitly using the intervals associated with bit-vectors. In another word, an allocation node *an* with absolute number *index* is added to bit-vector $v$ if $index \geq v.lower$ and $index \leq v.upper$ where *v.lower* and *v.upper* are lower and upper bounds of the interval associated with $v$. In real implementation of our method this assumption is relaxed which will be explained later in this section.

Each **RangedBitVector** consists of an array of chunks. For example 64-bit chunks (i.e. long[]) or 32-bit chunks (i.e. int[]) in current JVMs. The **or** operation is implemented by **or**ing corresponding chunks of each bit-vector. To implement **RangedBitVector**'s **or** operation, we cannot simply **or** the $i$th chunk of the first bit-vector and $i$th chunk of the second bit-vector since each bit in a **RangedBitVector** must be interpreted based on its lower bound. Figure 3 shows **RangedBitVector**'s **or** operation as we have implemented. Any two intervals created in the procedure shown in Figure 1 are disjoint, or one of them is subrange of the other one, so only these cases are handled in Figure 3. Variable declarations are omitted for saving space. The constant CHUNK_SIZE shows the size of each chunk in the implementation of **RangedBitVector** (32 bits in the previous example).

The overall goal of the procedure shown in Figure 3 is to add all the members of **RangedBitVector** y which fall within the interval associated with x, to x and return true if x is changed and false otherwise.

The procedure is conservative in that it aligns the portion which is common to the intervals associated with **RangedBitVector**s x and y so it can do the **or** operation chunk-by-chunk (instead of bit-by-bit). This relaxes the assumption we made earlier in this section and as a consequence reduces the precision of type filtering compared to the type filtering performed by using type masking during the propagation. We found this reduction of precision trivial according to the experiment we did. See section VII.

## VI. Ranged points-to set and hybrid ranged points-to set

Mainly, two versions of points-to sets have been added to Spark. One of them is **RangedPointsToSet** which has a **RangedBitVector** for every interval that is associated with its type. Our second major set implementation that is faster and more memory efficient than the first one is inspired from Spark's hybrid points-to set. It contains 16 allocation node references which are initialized to null, whenever all of them are filled up and the set is going to get larger than 16 members, it will become a **RangedPointsToSet**.

## VII. Experimental results

We have incorporated our techniques into Spark which is a research framework for points-to analysis. We observed that our work improves the memory allocated to points-to sets by a factor of about 2.5 compared to Spark's hybrid points-to set. It also improves the propagation time for sufficiently large programs compared to Spark's hybrid points-to set which is the fastest and default points-to set in current version of Soot [19] (version 2.4.0).

There is an implementation of Hientze's shared points-to set in Spark which consumes less memory than hybrid's points-to set and our implementation, but it is slower than both of them. See Table I. The results which are shown for Heintze's sets may be smaller than what they really consume since we calculate the results after the propagation but it may take more memory at some point during the propagation.

To evaluate our work we chose 4 programs as follows, jEdit 2.4 (a text editor), JFlex 1.4.3 and SableCC 2.18.2 (both of them are lexical analyzer generators) and Soot 1.2.5 (a framework for analyzing and optimizing Java bytecode).
We used JRE 1.3 as the library to analyze the input programs. All benchmarks are performed on a machine with 2 GB memory (1 GB allocated to JVM) and 2 GHz Intel core 2 Duo CPU running Ubuntu 8.10.

| program | hybrid | | range+hybrid | | heintze | |
|---|---|---|---|---|---|---|
| | time | space | time | space | time | space |
| jedit | 12.5 | 79.5 | 8.5 | 31.9 | 18.6 | 4.9 |
| soot | 10.0 | 98.4 | 8.5 | 34.4 | 28.1 | 5.4 |
| jflex | 5.0 | 65.2 | 5.3 | 25.9 | 13.2 | 4.1 |
| sablecc | 1.6 | 17.2 | 2.0 | 8.8 | 3.1 | 1.5 |

TABLE I
TIME AND SPACE FOR POINTS-TO SET PROPAGATION (SPACE IN MB AND TIME IN SECONDS). COMPARISON BETWEEN SPARK'S HYBRID, OUR RANGED-HYBRID AND HEINTZE'S SHARED POINTS-TO SETS. THE COLUMN RANGE+HYBRID SHOWS OUR IMPLEMENTATION.

Programs jEdit, Soot, JFlex and SableCC consist of 13583, 13741, 11893 and 5299 methods respectively. This shows that our most efficient points-to set (**HybridRangedPointsToSet**) makes the analysis of the two larger programs faster (JEdit and JFlex) and consumes significantly less memory in all four cases when compared to Spark's hybrid set (the fastest and default points-to set of Spark in current version of Soot).

We also compared our implementation to sparse bitmaps [16], [17] in term of space that would be saved if our techniques were combined with the sparse bitmaps (i.e. every element in a sparse bitmap would be a **RangedBitVector** of size $8 \times 4$ bytes and no element would be allocated unless it has at least one bit set). Table II shows the results of this experiment. Considering the results shown in the table and the cost related to the linked list manipulation of elements in sparse bit-maps, we believe that sparse bitmaps are not a really good candidate to be combined with our techniques. One reason could be that the **RangedBitVector**s are not sparse because of the way we numbered allocation nodes.
Since in our ordering, all allocation nodes of compatible types are consecutive and a points-to set of type T is filled only with those of compatible type with T, but in the plain sparse method, allocation nodes of type T would be scattered and are not grouped in some interval. Hence the sparse method combined with our way of numbering has higher chance to allocate fewer elements.

To compare the precision of our intrinsic type filtering with type filtering based on type masking, we considered all variable nodes which are dereferenced. Table III shows the percentage of points-to sets with 0, 1, 2, 3 - 10, 11 - 100, 101 - 1000 and more than 1000 elements in their points-to sets. According to this table, reduction of precision which is caused by the way we do type filtering (alignment of the common subrange as described in section V) is trivial.

| program | range | range+hybrid |
|---------|-------|--------------|
| jedit | 49.3/6.1 | 31.9/6.4 |
| soot | 44.7/5.9 | 34.4/5.9 |
| jflex | 37.3/4.6 | 25.9/4.8 |
| sablecc | 9.6/1.1 | 8.8/1.1 |

TABLE II
THIS TABLE SHOWS THE SPACE THAT WOULD BE SAVED IF OUR TECHNIQUES WERE COMBINED WITH SPARSE BITMAPS. THE COLUMN RANGE SHOWS **RANGEDPOINTSTOSET** AS DESCRIBED IN SECTION VI AND THE SECOND COLUMN SHOWS OUR MOST EFFICIENT SET IMPLEMENTATION AS DESCRIBED IN SECTION VI (TOTAL SET SIZE / SPACE THAT WOULD BE SAVED IF WAS USED IN COMBINATION WITH SPARSE BITMAPS).

|  | jedit | jflex | soot | sablecc |
|---|-------|-------|------|---------|
| 0 | 5.48/5.48 | 4.49/4.49 | 1.41/1.41 | 6.37/6.37 |
| 1 | 28.99/29.01 | 31.70/31.72 | 37.85/37.87 | 34.99/34.99 |
| 2 | 7.11/7.12 | 6.89/6.91 | 17.90/17.90 | 10.72/10.80 |
| 3-10 | 47.99/48.02 | 47.00/47.05 | 36.17/36.19 | 39.95/39.95 |
| 11-100 | 7.81/7.87 | 9.47/9.39 | 5.28/5.28 | 7.71/7.63 |
| 101-1000 | 2.53/2.42 | 0.35/0.35 | 1.29/1.26 | 0.22/0.22 |
| 1000+ | 0.0/0.0 | 0.0/0.0 | 0.0/0.0 | 0.0/0.0 |

TABLE III
PRECISION OF INTRINSIC TYPE FILTERING AND TYPE FILTERING BASED ON TYPE MASKING - PRECISION: INTRINSIC (**RANGEDPOINTSTOSET**) / TYPE MASKING (PURE BIT-VECTOR) (% OF TOTAL)

## VIII. CONCLUSION

In this paper, we presented and evaluated an improvement to points-to analyses which orders allocation nodes in our desired way and uses our implementation of points-to set. This work also combined our methods with Spark's hybrid points-to set idea and evaluated the performance gain.

Type filtering is done in our method implicitly (it is done by means of the **RangedBitVector**s). In another word, we do not need the additional logical **and** operation to enforce type filtering (type masks). This feature along with having smaller bit-vectors made our method more efficient than the previous works. Note that to do the **or** operation in the units of chunk some justification is performed as you saw in Figure 3 which is an additional overhead compared to previous works. Despite this overhead, we see an improvement in time for sufficiently large programs. See section VII.

We observed that our work consumes $2.5\times$ less memory than hybrid points-to set (default set in Spark). It also improves points-to propagation time for sufficiently large programs compared to the same set. We also compared our set implementation to other state of the art implementations including Hientze's shared bit-vectors [7] and sparse bitmaps [16], [17]. We observed that Hientze's shared bit-vector sets use less memory than both our implementation and Spark's hybrid sets but it is also slower than both of them. We did not find sparse bit-maps a good candidate to be combined with our method and we believe that it is practically less efficient than our method (see section VII).

Finally, you can find further details like the way allocation nodes of array types are handled in our technical report [18].

ACKNOWLEDGMENT

We would like to thank Ondrej Lhotak, Laurie Hendren and other members of Sable research group for developing Soot and making it available on-line.

We are also grateful to other members of IPLP research group, Yaser Elmi and Saeed Paktinat, who have helped us in understanding the points-to analysis problem.

REFERENCES

[1] Lars Ole Andersen. Program analysis and specialization for the C programming language. PhD thesis, University of Copenhagen, 1994.
[2] John Whaley and Monica Lam. Clonning-based context-sensitive pointer alias analysis using Binary Decision Diagrams. In Programming Language Design and Implementation (PLDI), 2004
[3] Ondrej Lhotak and Laurie Hendren. Scaling Java points-to analysis using Spark. In Compiler Construction, 12th International Conference, volume 2622 of LNCS, Springer, 2003
[4] Ondrej Lhotak, Stephen Curial, and Jose Nelson Amaral. Using ZBDDs in points-to analysis. In Languages and Compilers for Parallel Computing, 20th International Workshop, LCPC, 2007.
[5] Jianwen Zhu. Symbolic pointer analysis. In International Conference on Computer Aided Design (ICCAD), November 2002.
[6] Marc Berndl, Ondrej Lhotak, Feng Qian, Laurie Hendren, and Navindra Umanee. Points-to analysis using BDDs. In Proceedings of the ACM SIGPLAN 2003 Conference on Programming Language Design and Implementation, pages 103-114. ACM Press, 2003.
[7] Nevin Hientze. Analysis of large code bases, the compile-link-analyze model. Draft version, November 1999.
[8] Martin Hirzel, Daniel Von Dincklage, and Amer Diwan. Fast online pointer analysis. ACMTransactions on Programming Languages and Systems, April 2007.
[9] Ondrej Lhotak, Stephen Curial, and Jose Nelson Amaral. Using XBDDs and ZBDDs in points-to analysis. Software, Practice and Experience, 39(2):163-188, 2009.
[10] Jerey Dean, David Grove, and Craig Chambers. Optimization of object oriented programs using static class hierarchy analysis. In ECOOP 95, August 1995.
[11] David F. Bacon and Peter F. Sweeney. Fast static analysis of C++ virtual function calls. In OOPSLA'96: Conference Proceedings: Object-Oriented Programming Systems, Languages, and Applications, pages 324-341, 1996.
[12] Vijay Sundaresan, Laurie J. Hendren, Chrislain Razafimahefa, Raja Vallee-Rai, Patrick Lam, Etienne Gagnon, and Charles Godin. Practical virtual method call resolution for java. In Conference on Object-Oriented Programming, Systems, Languages, and Applications (OOPSLA '00), pages 264-280, 2000.
[13] Barbara Ryder. Dimensions of precision in reference analysis of object oriented languages. In International Conference on Compiler Construction (CC), 2003.
[14] James Gosling, Bill Joy and, Gilad Bracha. The Java language specification -third edition, 2005.
[15] Calvin Lin and Ben Hardekopf. Semi-Sparse Flow-Sensitive Pointer Analysis. Symposium on Principles of Programming Languages, 2009, pp. 226-238.
[16] GCC, the GNU Compiler Collection. http://gcc.gnu.org
[17] LLVM, The LLVM Compiler Infrastructure Project. http://llvm.org
[18] Hamid A. Toussi and Ahmed Khademzadeh. IPLP Resesrch Group Technical Report: Improving bit-vector representation of points-to sets using class hierarchy analysis. Islamic Azad University of Mashhad, July 2010.
[19] Vijay Sundaresan, Patrick Lam, Etienne Gagnon, Raja Vallee-Rai, Laurie Hendren and, Phong Co. Soot - a Java optimization framework. In Proceedings of CASCON 1999, pages 125-135, 1999.